\documentclass[equations,11pt]{article}\relax
\usepackage{amssymb,amsmath,amsthm} \relax
\usepackage{amsmath}
\usepackage{eucal}

\renewcommand{\theequation}{\arabic{equation}}
\begin{document}
\bibliographystyle{plain}
\def\m@th{\mathsurround=0pt}
\mathchardef\bracell="0365 
\def\upbrall{$\m@th\bracell$}
\def\undertilde#1{\mathop{\vtop{\ialign{##\crcr
    $\hfil\displaystyle{#1}\hfil$\crcr
     \noalign
     {\kern1.5pt\nointerlineskip}
     \upbrall\crcr\noalign{\kern1pt
   }}}}\limits}
\def\theequation{\arabic{section}.\arabic{equation}}

\newcommand{\pii}{P$_{{\rm\small II}}$} 
\newcommand{\pvi}{P$_{{\rm\small VI}}$}  
\newcommand{\ptf}{P$_{{\rm\small XXXIV}}$}  
\newcommand{\diffE}{O$\triangle$E} 

\newcommand{\pp}{\partial}
\newcommand{\ar}{\alpha}
\newcommand{\aar}{\bar{a}}
\newcommand{\bb}{\beta}
\newcommand{\gm}{\gamma}
\newcommand{\Gm}{\Gamma}
\newcommand{\en}{\epsilon}
\newcommand{\ven}{\varepsilon}
\newcommand{\dd}{\delta}
\newcommand{\sg}{\sigma}
\newcommand{\kp}{\kappa}
\newcommand{\ld}{\lambda}
\newcommand{\bmu}{\bar{\mu}}
\newcommand{\vf}{\varphi}
\newcommand{\Ups}{\Upsilon}
\newcommand{\oa}{\omega}
\newcommand{\hf}{\frac{1}{2}}
\newcommand{\bea}{\begin{eqnarray}}
\newcommand{\eea}{\end{eqnarray}}
\newcommand{\bse}{\begin{subequations}}
\newcommand{\ese}{\end{subequations}}
\newcommand{\nn}{\nonumber}
\newcommand{\bR}{\bar{R}}
\newcommand{\bP}{\bar{\Phi}}
\newcommand{\bS}{\bar{S}}
\newcommand{\bu}{{\boldsymbol u}}
\newcommand{\bt}{{\boldsymbol t}}
\newcommand{\bm}{{\boldsymbol m}}
\newcommand{\boa}{{\boldsymbol \omega}}
\newcommand{\bet}{{\boldsymbol \eta}}
\newcommand{\bW}{\bar{W}}
\newcommand{\sn}{{\rm sn}}
\newcommand{\wh}{\widehat}
\newcommand{\ol}{\overline}
\newcommand{\wt}{\widetilde}
\newcommand{\ut}{\undertilde}
 \newcommand{\bU}{\bf U}
 \newcommand{\pl}{\partial}
 \newcommand{\ddp}{\frac{\partial}{\partial p}}
 \newcommand{\ddq}{\frac{\partial}{\partial q}}
 \newcommand{\ddr}{\frac{\partial}{\partial r}}
 \newcommand{\Ld}{{\bf \Lambda}}
 \newcommand{\tLd}{\,^{t\!}{\bf \Lambda}}
 \newcommand{\I}{{\bf I}}
 \newcommand{\tII}{\,^{t\!}{\bf I}}
 \newcommand{\tuk}{\,^{t\!}{\bf u}_k}
 \newcommand{\tul}{\,^{t\!}{\bf u}_\ell}
 \newcommand{\tcl}{\,^{t\!}{\bf c}_{\ell}}
 \newcommand{\ssk}{\sigma_{k^\prime}}
 \newcommand{\ssl}{\sigma_{\ell^\prime}}
 \newcommand{\ddint}{\int_\Gamma d\ld(\ell) }
 \def\hypotilde#1#2{\vrule depth #1 pt width 0pt{\smash{{\mathop{#2}
 \limits_{\displaystyle\widetilde{}}}}}}
 \def\hypohat#1#2{\vrule depth #1 pt width 0pt{\smash{{\mathop{#2}
 \limits_{\displaystyle\widehat{}}}}}}
 \def\hypo#1#2{\vrule depth #1 pt width 0pt{\smash{{\mathop{#2}
 \limits_{\displaystyle{}}}}}}
 
\newtheorem{theorem}{Theorem}[section]
\newtheorem{lemma}{Lemma}[section]
\newtheorem{cor}{Corollary}[section]
\newtheorem{prop}{Proposition}[section]
\newtheorem{definition}{Definition}[section]
\newtheorem{conj}{Conjecture}[section]
 
\begin{center}
{\large{\bf On the discrete and continuous Miura Chain associated with 
the Sixth Painlev\'e Equation}}   
\vspace{1.2cm}
 
Frank Nijhoff\vspace{.15cm} \\
{\it Department of Applied Mathematics\\
The University of Leeds, Leeds LS2 9JT, UK}\\
\vspace{.5cm}
Nalini Joshi and Andrew Hone \vspace{.15cm}\\
{\it Department of Pure Mathematics\\
The University of Adelaide\\
Adelaide, Australia 5005}
\vspace{.5cm}

\today
\end{center}
\vspace{1.4cm}
 
\centerline{\bf Abstract}  
\vspace{.2cm}
A Miura chain is a (closed) sequence of differential (or difference)
equations that are related by Miura or B\"acklund transformations. 
We describe such a chain for the sixth Painlev\'e equation (\pvi ), 
containing, apart from \pvi\ itself, a Schwarzian version 
as well as a second-order second-degree ordinary differential 
equation (ODE). As a byproduct we derive
an auto-B\"acklund transformation, relating two copies of \pvi\  with
different parameters. We also establish the 
analogous ordinary difference equations in the discrete counterpart 
of the chain. Such difference equations govern iterations
of solutions of \pvi\ under B\"acklund transformations.                 
Both discrete and continuous equations constitute a larger system
which include partial difference equations, differential-difference 
equations and partial differential equations, all associated with 
the lattice Korteweg-de Vries equation subject to 
similarity constraints.  

\vfill

\setcounter{page}{0}
\pagebreak
 
\section{Introduction}  
\setcounter{equation}{0} 
The six Painlev\'e equations are second-order nonlinear 
ordinary differential equations (ODEs) that arose in the classification
programme originated by Painlev\'e \cite{Pain,Gambier}. In this paper,
we concentrate on the sixth Painlev\'e equation (first found by R. Fuchs in
\cite{Fuchs}):
\begin{equation*}
\phantom{aaaaaaa}\begin{array}{ll}
\displaystyle\frac{d^2w}{dt^2}&\displaystyle=\frac{1}{2}\left(
\frac{1}{w}+\frac{1}{w-1}+
\frac{1}{w-t}\right)\left(\frac{dw}{dt}\right)^2 -
\left(\frac{1}{t}+\frac{1}{t-1}+\frac{1}{w-t}\right)
\frac{dw}{dt}\\ &\\
&\displaystyle\ + \frac{w(w-1)(w-t)}{8t^2(t-1)^2}\left( \ar-\bb\frac{t}{w^2}
+\gm\frac{t-1}{(w-1)^2}-(\dd-4)\frac{t(t-1)}{(w-t)^2} \right),\\
\end{array}\ \quad
{\rm P}_{{\rm\small VI}}
\end{equation*}
and equations related to it by Miura or B\"acklund transformations.

Painlev\'e's classification programme aims
to identify classes of ODEs with the property that all
movable singularities of all solutions are poles. He and his school
completed
the classification work for first-degree second-order ODEs (under some
natural
assumptions). However, the work for third-order ODEs remains incomplete. The
third-order classification was undertaken by Chazy,
Garnier, and Bureau \cite{Chazy,Garnier} and more recently by Cosgrove
\cite{Cosgrove}. 
However, they restricted their attention to special classes which exclude
{\em Schwarzian}\/ equations \cite{Laine}, i.e. ODEs of the form
\begin{equation}\label{eq:Schwarzian}
\Bigl(\{Z, x\}\Bigr)^k=R(Z_{xx},Z_x,Z,x),\quad
\{Z,x\}:=
\frac{ Z_{xxx}}{Z_x}-\frac{3}{2}\frac{Z_{xx}^2}{Z_x^2}, 
\end{equation}
where $k$ is a positive integer and $R$ is a rational function of $Z_{xx}$,
$Z_{x}$, and
$Z$ with polynomial coefficients. Equations of Schwarzian form appear
naturally as
similarity reductions of soliton equations 
(just as the Painlev\'e equations
do \cite{ARS}).  For example, the third-order ODE 
\begin{equation}\label{eq:SPII}
Z'''-\frac{3}{2}\frac{{Z''}^2}{Z'}=\mu Z - \frac{1}{3}xZ'
\end{equation}
(where primes represent derivatives w.r.t. $x$) 
arises as a reduction of the Schwarzian KdV
equation
\begin{equation}\label{eq:uSKdV} 
\frac{z_t}{z_x}=\{ z,x \}  , 
\end{equation}
cf. \cite{weiss:II,Dorf}.                             
It is related to the second Painlev\'e 
equation 
\begin{equation*}
\phantom{aaaaaaaaaaaaaaaaaaaaaaa}V''=2V^3-\frac{1}{3}xV-\frac{3\mu-1}{6}   
\qquad\qquad\qquad\qquad\qquad\qquad
{\rm P}_{{\rm\small II}}
\end{equation*}
by the Cole-Hopf transformation ~$V=-Z''/(2Z')$~. Even though equations of
the type 
(\ref{eq:Schwarzian}) are 
not M\"obius invariant (whereas the Schwarzian operator is),     
the partial differential equations (PDEs) from which they arise as
similarity
reductions {\em are} invariant under the action of $PSL_2(\mathbb{C})$.
Moreover,
they appear to possess some special symmetries that give rise to
B\"acklund transformations of the Painlev\'e equations to which they are
related.

In this note we present a generalization of Eq (\ref{eq:SPII}),  
namely a four-parameter Schwarzian \pvi\ equation (S\pvi ) (see Eq
(\ref{eq:SPVI}) below)
and the Miura transformation relating it to \pvi . In addition, we give a
second-order
second-degree equation (see Eq. (\ref{eq:hh}) below)       
which we call a {\em modified} \pvi\ (M\pvi
) which is also related to \pvi\ by a Miura transformation. As a byproduct,
we obtain an auto-B\"acklund transformation for \pvi\ (i.e. a transformation
relating \pvi\ to a copy of itself with possibly different parameters). 
These three transformations form what we call the {\em Miura chain}. 

It should be noted that relationships  
between \pvi\ and second-degree equations were known before, cf. 
\cite{Bureau,FY,FokAb,CosScou}. However, our point of view that the 
latter equation constitutes  a modified version of \pvi\ appears to
be new, and is motivated by the connection of these equations 
to certain discrete equations which forms the central part of the 
present note. In fact, 
we present the exact discrete counterparts of the S\pvi\ and
M\pvi\  equations. These are nonlinear nonautonomous ordinary difference
equations that 
can be considered to be nonlinear superposition formulae for the 
auto-B\"acklund transformations (auto-BTs) of the corresponding continuous 
equations. In recent years such difference equations have attracted 
a great deal of attention (see \cite{Carg} for a review) as integrable
difference equations, possibly defining new transcendental 
functions by means of discrete rather than continuous equations. 

The derivation of the equations is based on the results on the similarity 
reduction of a family of partial difference equations, namely the KdV 
family of lattice systems, a programme that was initiated in Ref. 
\cite{NP} and continued in \cite{DIGP}. The present note is a further
development of the general framework presented in a       
recent paper \cite{NRGO},
in which equations on the lattice as well as in the continuum 
are naturally linked together. It was shown there that      
\pvi\ (with general parameters $\ar$, $\bb$, $\gm$, $\dd$)                
naturally arises from such a framework for the lattice
KdV family (without imposing a limit). Moreover, a new              
discrete Painlev\'e equation with four free parameters
was derived, the general solution of which can be expressed in         
terms of the transcendental solutions of \pvi . We will refer to these 
continuous and discrete results as forming the \lq\lq regular\rq\rq\   
\pvi\ system.                   

In the present paper we extend                 
the results of \cite{NRGO}. We derive S\pvi , M\pvi\ as 
well as their natural discrete analogues, and transformations between
them.
Thus we establish a Miura chain also for 
ordinary difference  equations (\diffE s). As is the case
with  the regular \pvi\ system, these \diffE s govern the same functions as
the
corresponding ODEs and are compatible with them.  Hence
there are common solutions to both the discrete and
continuous equations. 
In our view, these results might shed some new light on the algebraic 
structures behind solutions of \pvi , cf. e.g. 
\cite{Oka}. 

The Miura chain can be schematically described as in the following      
diagram.
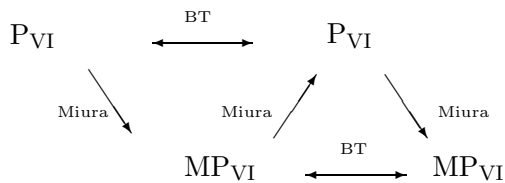
\begin{center}      
\begin{picture}(400,50)(-120,5)
\put(0,50){\pvi }
\put(60,50){\vector(1,0){32}}\put(60,50){\vector(-1,0){6}}
\put(120,50){\pvi }
\put(30,40){\vector(2,-3){16}}
\put(18,22){{\tiny Miura}}
\put(80,22){{\tiny Miura}}
\put(100,14){\vector(2,3){16}}
\put(162,22){{\tiny Miura}}
\put(142,38){\vector(2,-3){16}}
\put(66,58){{\tiny BT}}
\put(125,8){{\tiny BT}} 
\put(66,0){M\pvi }
\put(118,0){\vector(1,0){32}}\put(118,0){\vector(-1,0){6}}
\put(160,0){M\pvi }
\end{picture}
\end{center}
Note that the pull-back of the Miura 
between the modified and regular \pvi\ equations gives rise to 
an auto-BT. Behind this diagram there is another one for the corresponding
discrete
equations regulating the shifts in the parameters
$n$,$m$, $\mu$ and $\nu$ of the lattice KdV system (see e.g. Eq
(\ref{eq:MKdVsyst})) and
associated similarity constraints (see Eq (\ref{eq:MKdVconstr})). 

The outline of the paper is 
as follows: in \S 2 we present our results             
on three classes of equations: the Schwarzian \pvi\ class, the modified 
\pvi\ class and the regular \pvi\ class. 
We also give the formulae relating the various classes. 
In \S 3 we outline the derivation of the equations. Only an outline is
given here because more complete explanations of the lattice framework can
be found in \cite{NRGO}. In \S 4 we
demonstrate some  special parameter limits of the equations, some of which
have been 
given before in the literature. Finally in \S 5 we give some
concluding remarks. 
 
\section{The Painlev\'e VI Miura Chain} 
\setcounter{equation}{0}

In this section, we list the three classes of systems   
i.e. Schwarzian, Modified, Regular, which are intertwined through
the Miura chain. Both continuous and discrete equations are given.
An outline of their derivation may be found in \S 3.

We list them in the order given schematically as
\vspace{.1cm}
\begin{center}
{\sf Schwarzian \pvi }\hspace{.3cm} $\longrightarrow$ \hspace{.3cm}
{\sf Modified \pvi }\hspace{.3cm} $\longrightarrow$ \hspace{.3cm}
{\sf \pvi }\hspace{.3cm}
\end{center}
 \noindent
For each system, there are associated or component systems that can be
described
as follows:
\begin{enumerate}
\item The {\em lattice equation}, i.e. the partial difference equation 
governing the dependent variable, e.g. $z_{n,m}$, (where $n$ and $m$ are 
the lattice variables, whilst there are also lattice {\it parameters},  
e.g. $p$,$q$, or alternatively $r=q/p$ appearing in the equation); 
\item The {\em similarity constraints} compatible with the corresponding
lattice 
equation \cite{NRGO}, containing additional parameters $\mu$, $\nu$; 
\item The {\em ordinary difference equation} for the dependent variable 
obtained by using similarity constraints and lattice equations to eliminate 
shifts in one of $m$ or $n$ (thus leading to \diffE s governing a function
of one
lattice variable e.g. $z_n:=z_{n,m}$);
\item The {\em differential-difference equation}\/ for the dependent
variable as a function of a lattice variable, e.g. $n$, and a continuous
parameter, e.g. $r$;
\item The {\em ordinary differential equation} that results from the 
elimination of the shifts in the lattice variables altogether by 
usage of the similarity contraints in favour of
continuous operations only.  
\end{enumerate}
It should be noted that no continuum limit is carried out here. All
equations
govern the same functions as in the original lattice equations, subject to
the accompanying similarity constraints. 

\subsection{Schwarzian Painlev\'e VI System}  
The Schwarzian equations stand at the base of the Miura chain and 
seem to be the most fundamental in the family. We use them as the 
starting point for the derivations of other members in the 
\pvi\  family of equations. The relevant equations are listed below. 
\begin{itemize}
\item \underline{\sf Lattice Equation}
\bse\label{eq:SKdV}\begin{equation}\label{eq:dSKdV}
\frac{(z_{n,m}-z_{n+1,m})(z_{n,m+1}-z_{n+1,m+1})}{(z_{n,m}-z_{n,m+1})
(z_{n+1,m}-z_{n+1,m+1})} = r^2    
\end{equation}
see Ref. \cite{KDV}.                                          
\item \underline{\sf Similarity Constraint \# 1}
\begin{equation} \label{eq:skdvconstr1}
n\frac{(z_{n+1,m}-z_{n,m})(z_{n,m}-z_{n-1,m})}{ z_{n+1,m}-z_{n-1,m}}
+ m \frac{ (z_{n,m+1}-z_{n,m})(z_{n,m}-z_{n,m-1})}{z_{n,m+1}-
z_{n,m-1} } = \bmu z_{n,m}\   ,
\end{equation}
see Ref. \cite{Dorf}, where we use the notation $\bmu=\mu+\frac{1}{2}$  
to remain consistent with that of \cite{NRGO}. 
\item \underline{\sf Similarity Constraint \# 2}
\begin{equation}
\label{eq:skdvconstr2}
\mu-\nu=n\frac{z_{n+1,m}+z_{n-1,m}-2z_{n,m}}
{ z_{n+1,m} - z_{n-1,m} }
+m\frac{z_{n,m+1}+z_{n,m-1}-2z_{n,m}}{ z_{n,m+1} - z_{n,m-1} }\   ,
\end{equation}
in which 
$\nu=\ld(-1)^{n+m}$ is an alternating coefficient with $\ld$ constant.  
\item \underline{\sf Ordinary Difference Equation}
\bea\label{eq:dSPVI} 
&&(r^2-1)(z_{n+1}-z_n)^2  = \nn\\ 
&&= \left[ 2r^2\frac{\bmu z_{n+1}(z_{n+2}-z_n)-(n+1)
(z_{n+2}-z_{n+1})(z_{n+1}-z_n)}{(m-\mu-\nu)(z_{n+2}-z_n)+(n+1)
(z_{n+2}-2z_{n+1}+z_n)}+z_{n+1}-z_n\right] \times \nn\\ 
&&~~~~\times\left[ 2r^2\frac{\bmu z_n(z_{n+1}-z_{n-1})-n
(z_{n+1}-z_n)(z_n-z_{n-1})}{(m-\mu+\nu)(z_{n+1}-z_{n-1})+n
(z_{n+1}-2z_n+z_{n-1})}+z_n-z_{n+1}\right] \   ,  \nn\\ 
\eea
i.e. the discrete Schwarzian PVI equation. 
\item \underline{\sf Differential-Difference Equation}
\begin{equation} \label{eq:skdvdif}
-t\frac{d z_n}{dt}=n\frac{(z_{n+1}-z_n)(z_n-z_{n-1})}{z_{n+1}-z_{n-1}}
\end{equation}
\item \underline{\sf Ordinary Differential Equation}
\begin{eqnarray} \label{eq:SPVI} 
&& \{z,t\}= \Big[(a-b)^2\bmu^4 z^4 - 2(a-b)^2(t-2)t\bmu^3 z^3z^\prime  
+ \nn \\ 
&& ~~~ +(t-1) t^4 {z^\prime}^3 
\Big( -[(a-b)^2-4(1-t)^2-4at+2abt-2b^2t-4ct+2(a+b)ct+ \nn \\
&& (b-c)^2t^2-16\bmu + 32t\bmu-8t^2\bmu-12\bmu^2+12t\bmu^2-4t^2\bmu^2] 
z^\prime+ \nn \\
&& 8 t (2-3t+t^2) \bmu z^{\prime\prime}\Big)+\nn  \\
&& t^2\bmu^2 z^2 \Big( [4+6(a-b)^2-16t-4at-5a^2t+12abt- \nn \\
&& 7b^2t-4ct+2(a+b)ct+12t^2+4at^2-2abt^2+ \nn \\
&& b^2t^2+4ct^2-2act^2-c^2t^2+4(t-1)\bmu^2] {z^\prime}^2+ \nn \\
&& 8t(2-5t+3t^2) z^{\prime}z^{\prime\prime}+8(t-1)^2t^2 
   {z^{\prime\prime}}^2\Big)- \nn \\
&& 4(t-1)t^3\bmu z z^{\prime} \big( ( (a-b)^2-2t-2at+(a-b)bt- \nn \\
&& 2ct+act+bct-4\bmu+6t\bmu-4{\bmu }^2+2t\bmu^2) {z^{\prime}}^2+ \nn \\
&& 2t(2-4t+t^2-2\bmu+2t\bmu) z^{\prime} z^{\prime\prime}+ 
   t^2(t-1)(t-2){z^{\prime\prime}}^2\big)\Big]\Big/ \nn \\
&& \Big[8 {{(t-1)}^2} t^4 {{z^{\prime}}^2} 
(\bmu z+tz^{\prime}) (-\bmu z+(t-1)t z^{\prime})\Big]\   ,  
\end{eqnarray}
\ese
where the parameters $a$,$b$,$c$ are given by           
\[ 
a=\frac{1}{2}+\nu+m-n \         \ ,\      \
b=\frac{1}{2}+\nu+m+n \         \ ,\      \
c=\frac{1}{2}+\nu-m+n \   .
\] 
and the independent variable $t=1/r^2$. Eq (\ref{eq:SPVI}) is
the Schwarzian \pvi .
\end{itemize}

\subsection{Painlev\'e VI System} 
We now move on to the ``regular'' \pvi\ equations, i.e. the equations 
that give rise directly to \pvi\ itself. These are given in terms 
of the intermediate variable 
$v_{n,m}$ which is related to the Schwarzian variable $z_{n,m}$ 
via the (discrete Cole-Hopf type) relations 
\bse\label{eq:dHC}\bea
p(z_{n,m}-z_{n+1,m})&=&v_{n+1,m}v_{n,m}\   , \\
q(z_{n,m}-z_{n,m+1})&=&v_{n,m+1}v_{n,m}\   .
\eea\ese
The main objects of interests, however, are combinations of the 
variables $v_{n,m}$ given by
\begin{equation} \label{eq:x-y}  
x_{n,m}\equiv \frac{v_{n,m}}{v_{n+1,m+1}}\      \ ,\      \ 
y_{n,m}\equiv \frac{v_{n+1,m}}{v_{n,m+1}}\    ,   
\end{equation}
which are related through the fractional linear transformation 
\begin{equation} \label{eq:xy} x_{n,m}=\frac{y_{n,m}-r}{1-ry_{n,m}} 
\    \ \Leftrightarrow \    \
 y_{n,m}=\frac{x_{n,m}+r}{1+rx_{n,m}} \   .  \end{equation}
The following equations arise from the results of \cite{NRGO}: 
\begin{itemize}
\item \underline{\sf Lattice Equation}
\bse\label{eq:MKdVsyst}
\begin{equation}  \label{eq:dMKdV}
\frac{v_{n,m+1}}{v_{n+1,m+1}}-\frac{v_{n+1,m}}{v_{n,m}}
= r\left( \frac{v_{n+1,m}}{v_{n+1,m+1}}-\frac{v_{n,m+1}}{v_{n,m}}
\right) \   ,
\end{equation} 
\item \underline{\sf Similarity Constraint}
\begin{equation} \label{eq:MKdVconstr}
n\frac{v_{n+1,m}-v_{n-1,m}}{v_{n+1,m}+v_{n-1,m}} +
m\frac{v_{n,m+1}-v_{n,m-1}}{v_{n,m+1}+v_{n,m-1}}  = \mu-\nu \   ,
\end{equation}  
cf. \cite{NP}. 
\item \underline{\sf Ordinary Difference Equation}
\bea \label{eq:mpii}
&&(n+1)(r+x_n)(1+rx_n)
\frac{x_{n+1}-x_n+r(1-x_nx_{n+1})}{x_{n+1}+x_n+r(1+x_nx_{n+1})} \nn \\
&& ~~~~~~~~~~~~~~ -n(1-r^2)x_n
\frac{x_n-x_{n-1}+r(1-x_nx_{n-1})}{x_n+x_{n-1}+r(1+x_nx_{n-1})}\nn \\
&& ~~~ = \mu r(1+2rx_n+x_n^2)+\nu(r+2x_n+rx_n^2)
-mr(1-x_n^2),
\eea
cf. \cite{DIGP,NRGO}. The relation between the variable $x_n$ and 
$v_{n,m}$ is explained in section 3.  
\item \underline{\sf Differential-Difference Equation}
\begin{equation} \label{eq:mkdvdif}
-2t\frac{d}{dt}\log v_n=n\frac{v_{n+1}-v_{n-1}}{v_{n+1}+v_{n-1}}
\end{equation}
\item \underline{\sf Ordinary Differential Equation}
\bea\label{eq:altPVI}
&&(1-r^2)^2y(ry-1)(y-r)\left[\frac{2}{r}\frac{\pl y}{\pl r}- 
\frac{\pl ^2y}{\pl r^2}\right] =\nn \\ 
&&=\frac{1}{2}(1-r^2)^2\left[ r(3y^2+1)-2(1+r^2)y\right]
\left(\frac{\pl y}{\pl r}\right)^2 + \nn\\ 
&&~+(1-r^2)\frac{1}{r}\left[ 2y(y-r)(ry-1)+(1-r^2)^2y^2\right] 
\frac{\pl y}{\pl r}  \nn \\
&&~+\frac{1}{2r}\left[ (\ar y^2-\bb)(y-r)^2(ry-1)^2 \right. \nn \\ 
&& ~~~~~~~~~ \left. +(1-r^2)y^2
\left((\gm-1)(ry-1)^2-(\dd-1)(y-r)^2\right)\right]\  ,  
\nn \\ \eea \ese 
which is \pvi\ with the identifications    
$t=r^{-2}$, $w(t)=y/r$ and identifying the parameters 
\begin{eqnarray*} 
&&\ar=(\mu-\nu+m-n)^2 ~~~~~\         \ ,\      \
\bb=(\mu-\nu-m+n)^2\   ,  \nn \\
&&\gm=(\mu+\nu-m-n-1)^2\     \ ,\      \
\dd=(\mu+\nu+m+n+1)^2\   .
\end{eqnarray*} 
Note that the variables $n$, $m$ 
need not be integers: we only require that they {\it shift by units}. 
In (\ref{eq:altPVI}) the variable $y$ is fractionally linearly 
related to $x_n$ as is explained in section 3. 
\end{itemize}

\subsection{B\"acklund Transformation}  
It should be noted that the Schwarzian \pvi\ can be written in 
second-order form, namely: 
\bea\label{eq:newPVI}
&& 2t\frac{d}{dt}\left\{ \frac{t(1-t)W'-\frac{1}{2}[(a-b)+(c+b)t]W
+\frac{1}{2}\bmu t(c+b)}{(W-\bmu)[(t-1)W-\bmu t]}\right\}\nn\\ 
&& = \frac{cbt}{(1-t)W}-W -\frac{(c+b)\bmu t
[ t(1-t)W'-\frac{1}{2}[(a-b)+(c+b)t]W +\frac{1}{2}\bmu t(c+b) ]}
{(1-t)W(W-\bmu)[(t-1)W-\bmu t]} \nn \\ 
&&~~ + \left( W+\frac{\bmu^2t}{(1-t)W}\right) 
\frac{\left[t(1-t)W'-\frac{1}{2}[(a-b)+(c+b)t]W
+\frac{1}{2}\bmu t(c+b)\right]^2}{(W-\bmu)^2[(t-1)W-\bmu t]^2}\   ,  
\eea  
by using the transformation 
\begin{equation}\label{eq:zW}
W(t)=\bmu+t\frac{z'(t)}{z(t)}\   . 
\end{equation}
Since we know that \pvi\ is the only representative in the relevant 
class of the Painlev\'e-Gambier classification \cite{ince}, this 
second--order first-degree ODE must be equivalent to \pvi\ up to a 
M\"obius transformation.  
Indeed, eq. (\ref{eq:newPVI}) is the \pvi\ equation with the 
identifications:
\begin{equation}\label{eq:id}
W(t)=\bmu \bar{w}(\bar{t})\    \ ,\      \ \bar{t}=\frac{t}{t-1}\  , 
\end{equation}
where $\bar{w}(\bar{t})$ solves  \pvi\ with new parameters  
\begin{eqnarray*} 
&&\overline{\ar} = 4\bmu^2\                    \ , \    
       \ \overline{\bb}=(b-c)^2=4m^2   \nn \\ 
&&\overline{\gm} = (a-b)^2=4n^2\      \ , \    
   \ \overline{\dd}=(a+c-2)^2=(2\nu-1)^2\  .    
\end{eqnarray*} 
The connection with the modified \pvi\ equation takes place via 
a Miura type transformation giving rise to an auto-BT of the form: 
\begin{equation}\label{eq:Miura}
w=t-W^{-1}\left( \bmu+ 
(b-\bmu)\left[ 1- 
\frac{t(1-t)W'-\frac{1}{2}[(a-b)+(c+b)t]W
+\frac{1}{2}\bmu t(c+b)}{(W-\bmu)[(t-1)W-\bmu t]}
\right]^{-1} \ \right)    , 
\end{equation} 
for \pvi\ itself. Concrete auto-BT's for \pvi\ were given in e.g. 
\cite{Gromak,FY,FokAb}; however, this one seems to be different from those. 

The explicit relation between the Schwarzian equation (\ref{eq:SPVI}) 
and \pvi\ is given by (\ref{eq:Miura}) 
whereas the inverse relation is  
\begin{equation}\label{eq:invMiura}          
W=\left(1-\frac{w}{t}\right)^{-1}[\bmu+(b-\bmu)(1-h)^{-1}],  
\end{equation}
where in the above $h=h(w',w,t)$ is given by the 
formula (\ref{eq:cinvMiu}) below. 
Analogous connections also exist on the level of the O$\Delta$Es. 
In fact, as is explained in section 3, the relations between $x$, 
$y$ on the one hand and $z$ on the other hand can be written in the 
form
\begin{equation}\label{eq:ddHC} 
ry_{n,m}=\frac{z_{n+1,m}-z_{n,m}}{z_{n,m+1}-z_{n,m}}\     \ ,\     \ 
x_{n,m}=r\frac{z_{n,m+1}-z_{n,m}}{z_{n+1,m+1}-z_{n,m+1}}\    , 
\end{equation}
which reduces to a discrete Cole-Hopf transformation between the 
O$\Delta$Es (\ref{eq:dSPVI}) and (\ref{eq:mpii}) if we eliminate the 
shifts $z_{n,m}-z_{n,m+1}$ using the relation 
\begin{equation}\label{eq:zz} 
z_{n,m}-z_{n,m-1}=2\frac{\bmu z_{n,m}-nA_{n,m}}{\mu
-\nu+m-nD_{n,m}}\     \ ,\    \ 
z_{n,m+1}-z_{n,m}=2\frac{\bmu z_{n,m}-nA_{n,m}}{m-\mu
+\nu+nD_{n,m}}\   , 
\end{equation}
in which 
\[A_{n,m}\equiv \frac{(z_{n+1,m}-z_{n,m})(z_{n,m}-z_{n-1,m})}{(z_{n+1,m}
-z_{n-1,m})}\    \ ,\    \ 
D_{n,m}\equiv \frac{z_{n+1,m}+z_{n-1,m}-2z_{n,m}}{
z_{n+1,m}-z_{n-1,m}} .  \] 
Note that the latter objects only involve shifts in $n$. 

\subsection{Modified Painlev\'e VI System} 
The modified equations are formulated in terms of the canonical 
variables 
\begin{equation}  \label{eq:hk}
h_{n,m}\equiv\frac{z_{n+1,m}+z_{n,m}}{z_{n+1,m}-z_{n,m}}\     \ ,
\     \ 
k_{n,m}\equiv\frac{z_{n,m+1}+z_{n,m}}{z_{n,m+1}-z_{n,m}}\   .
\end{equation}
Using these expressions and rewriting the various relations 
in terms of the variables $h$ and $k$, we obtain the following 
set of equations.
\begin{itemize}
\item \underline{\sf Lattice Equations}
\bse\label{eq:KdVsyst}
\bea  \label{eq:dKdV}
&& k_{n+1,m}-k_{n,m} = r^2(h_{n,m+1}-h_{n,m})\   ,  \\ 
&& k_{n+1,m}k_{n,m}-1 = r^2(h_{n,m+1}h_{n,m}-1)\   ,  
\eea
\item \underline{\sf Similarity Constraint \# 1}
\begin{equation} \label{eq:dp34constr1}
\frac{2n}{h_{n-1,m}+h_{n,m}} +
\frac{2m}{k_{n,m-1}+k_{n,m}} = \bmu\   , 
\end{equation}  
\item \underline{\sf Similarity Constraint \# 2}
\begin{equation} \label{eq:dp34constr2}
\frac{2nh_{n,m}}{h_{n-1,m}+h_{n,m}} +
\frac{2mk_{n,m}}{k_{n,m-1}+k_{n,m}} = \frac{1}{2}+\nu+n+m
\   , \end{equation}  
\item \underline{\sf Ordinary Difference Equation}
\bea \label{eq:dP34}
&& (1-r^2) \left(\bmu-\frac{2(n+1)}{h_{n+1}+h_n}\right)
\left(\bmu-\frac{2n}{h_{n-1}+h_n}\right) \nn \\ 
&& ~~ = \frac{\left[ \bmu h_n-(\frac{1}{2}+\nu+n-m)\right] 
\left[ \bmu h_n-(\frac{1}{2}+\nu+n+m)\right] }{h_n^2-1}\  . 
\eea
\item \underline{\sf Differential-Difference Equation}
\begin{equation} \label{eq:kdvdif}
2t\frac{d h_n}{dt}= \left(\frac{2(n+1)}{h_{n+1}+h_n} - 
\frac{2n}{h_{n-1}+h_n}\right) (h_n^2-1)\   . 
\end{equation}
\item \underline{\sf Ordinary Differential Equation}
The modified \pvi\ equation can be written as the coupled system 
of first-order ODE's: 
\bea\label{eq:MPVI}
2t\frac{dk}{dt}&=& \frac{a-\bmu k}{1-t}(h-k) 
-\frac{b-\bmu k}{h-k}(k^2-1)\    ,  \label{eq:MPVIa}  \\  
2t\frac{dh}{dt}&=& \frac{c-\bmu h}{1-t^{-1}}(h-k) 
-\frac{b-\bmu h}{h-k}(h^2-1)\    ,  \label{eq:MPVIb} 
\eea  \ese
with the identifications of $a$, $b$, $c$  as before. Eliminating 
the variable $k(t)$, we are led to the following second-order 
second-degree equation for $h(t)$:                 
\bea \label{eq:hh} 
&&\,4t^2(t-1)^2 h''
\Bigl[ t(t-1)h''+(2t-1)h'+4\bmu^2h^3-3\bmu(b+c)h^2 
+2(bc-\bmu^2)h +\bmu(b+c)\Bigr] \nn\\
&&=t(t-1){h^\prime}^2 [a-b+(b+c-2)t+2\bmu(1-2t)h]\times\nn\\
&&\quad\qquad\qquad\times [a-b-2+(b+c+2)t+2\bmu(1-2t)h] \nn\\
&&\quad-2t(t-1)(2t-1)h'\Bigl[ 4\bmu^2h^3-3\bmu h^2(b+c) 
+2(bc-\bmu^2)h+(b+c)\bmu\Bigr]  \nn  \\ 
&&\quad+4\bmu^4 h^6-4\bmu^3 [1-a+c+2b+(2a+c-b-2)t]h^5 \nn\\
&&\quad+\bmu^2\Bigl[(1-a)^2+6b(1-a)+5b^2
+4(1-a)c+8bc+5(2a-2+c-b)(b+c)t-4\bmu^2\Bigr] h^4 \nn\\
&&\quad+\bmu\Bigl[ -(1-a)^2(b+c)+(1-a)\Bigl( 4\bmu^2(1-2t) + 
2t(b+c)^2-2b^2+2bc(4t-3)\Bigr) \nn \\ 
&& \quad\qquad\qquad + (4\bmu^2+bc)(b+c)t^2-b^3(t-1)^2 - 
c^3t^2 +8\bmu^2 b(1-t) +4c\bmu^2   \nn\\
&& \quad\qquad\qquad + (4t-5)b^2c-6bc^2t\Bigr] h^3 \nn \\ 
&&\quad+\Bigl[ 4\bmu^4t(1-t)+10(1-a)(b+c)\bmu^2t 
-14\bmu^2bct(t-1)-\bmu^2(6b+4c)(1-a)-8bc\bmu^2 \nn \\ 
&&\quad\qquad\qquad +(1-a)^2(bc-\bmu^2) 
-(3t-5)(t-1)\bmu^2b^2- (3t+2)\bmu^2tc^2+bc^3t^2 \nn \\ 
&&\quad\qquad\qquad + 2b^2c(1-a)-2bc^2t(1-a) 
+b^3c(1-t)^2+2b^2c^2t(1-t)-2cb^2t(1-a)\Bigr] h^2 \nn \\
&&\quad+\bmu\Bigl[ b^3(1-2t)+(c+b)^3t^2+6bc(1-a)(1-2t)+2b^2(1-t)
-2ab^2(1-t)-2c^2t(1-a) \nn \\ 
&&\quad\qquad\qquad+\Bigl((1-a)^2+4bct^2-4\bmu^2t(1-t)\Bigr)(b+c) 
+(5-12t)b^2c -2b^2ct \Bigr] h \nn \\  
&&\quad+t(1-t)\Bigl(\bmu^2(b+c)^2+2b^2c^2\Bigr)-(1-a)^2bc 
-(1-t)^2b^3c  \nn\\
&&\quad\qquad\qquad -t^2bc^3+2bc^2t(1-a)-2b^2c(1-t)(1-a)\   .  
\eea 
Second-order second-degree equations related to \pvi\ were given in 
e.g. \cite{Bureau,FY,FokAb,CosScou}, cf. also \cite{Oka}, however 
we have not established the precise connection between (\ref{eq:hh}) 
and those equations, even though in principle it should 
fit into the classification of \cite{Bureau,CosScou}. 
\end{itemize}

The various Miura relations that link the modified \pvi\ 
equations to the regular ones are given as follows.

\begin{itemize}
\item \underline{\sf Discrete Miura}
\bea  \label{eq:xh}
x_n&=&r^{-1}\frac{(1-r^2)+r^2h_n-k_n}{k_n-h_n}  \\
&=&\frac{\left[(\mu-\nu-n-m)+\bmu r^2(h_n-1)\right] 
(h_{n-1}+h_n)+2n(1-r^2)(h_n-1)}{r[n+m+\nu+\frac{1}{2}-\bmu  
h_n](h_n+h_{n-1})} \nn 
\eea
\item \underline{\sf Discrete Inverse Miura}
\begin{equation}  \label{eq:hx}
\frac{h_{n-1}+1}{h_n-1}=\frac{x_n+r}{(1+rx_n)x_{n-1}} 
\end{equation}
\item \underline{\sf Continuous Miura}
\bea  \label{eq:cMiu}
&& \left[ t(t-1)h'-(c-\bmu h)(1-h)(w-t)\right]^2 = \nn \\ 
&&\quad = t^2(t-1)^2{h'}^2 + t(t-1)(c-\bmu h)(b-\bmu h)(h^2-1)
\eea 
\item \underline{\sf Continuous Inverse Miura}
\bea \label{eq:cinvMiu}
2\bmu h &=& \frac{1}{1-w}\Bigl[-a+(2+b)(1-t)-2t(1-t)\frac{w'}{w}+ \bmu t 
(2-\frac{w}{t})  \nn \\ 
&&\quad\qquad -c(w-t) + (a-\bmu)\frac{t}{w}\Bigr] 
\eea 
\end{itemize}

This completes the list of the various continuous and discrete 
equations in the Miura chain associated with the \pvi\ equation. 
What is manifestly absent in the above list in each of the three 
classes is the corresponding partial differential equation 
(PDE). It is important to realise that the full PDE analogue of 
e.g. (\ref{eq:dSKdV}) can{\it not} be simply the Schwarzian KdV 
(\ref{eq:uSKdV}), since the latter equation is not rich enough to 
yield full \pvi\ as a similarity reduction. In fact, it turns out 
that there does indeed exist a PDE that is associated with the 
\pvi\ system, but this PDE being a very rich 
equation in its own right, we believe it merits a separate 
publication, \cite{SPDE}.

\section{Remarks on the Derivations} 
\setcounter{equation}{0}

In this section, we outline the derivation of the results given   
in \S 2. Full details of the general framework within which we work 
were presented in Ref. \cite{NRGO}. We will not go into the details
of this structure here. However, it is worthwhile to note that there
is a subtle duality between the set of lattice {\em parameters} $p,
q, r, \ldots$ and the lattice {\em variables}  $n,m$ in this framework. 
The words \lq\lq parameters\rq\rq , \lq\lq variables\rq\rq\ follow naturally
from the roles they play in the discrete framework. However, in the
continuous equations
we derive, the parameters play the role of independent variables, whereas
the lattice
variables play the role of parameters. So the shifts in the lattice 
variables have the interpretation of B\"acklund transformations 
for the continuous equations. 

The similarity constraints form the starting point for the derivation of
results in \S 2. In particular, the ones for the Schwarzian system
play an important role. In Ref. \cite{NRGO},
the simlarity constraint for the variables $v$, i.e. eq.
(\ref{eq:MKdVconstr}), 
were derived from one for the variable $z$, i.e. (\ref{eq:skdvconstr1}).
(This involves taking first a                       
difference in the discrete variables $n$ or $m$ and then using the 
P$\Delta$E (\ref{eq:dSKdV}) to rewrite it in a form that can integrated 
once, thus leading to a new alternating parameter $\nu$.) The resulting
constraint for $v$ again in terms of $z$ led to the second similarity
constraint
(\ref{eq:skdvconstr2}). Both the Schwarzian constraints are used to derive
the ones for $h$ and $k$, i.e. Eqs (\ref{eq:dp34constr1}) and
(\ref{eq:dp34constr2}).
The combination of the two constraints (\ref{eq:skdvconstr1}) and 
(\ref{eq:skdvconstr2}) lead to the relations (\ref{eq:zz}). The latter,
together with the
P$\triangle$E used to eliminate shifts in $m$, gives rise to the
\diffE\ (\ref{eq:dSPVI}).  

The derivation of \pvi\ in \cite{NRGO} uses two additional intermediate 
variables:
\begin{equation} \label{eq:a-b}
a_{n,m}\equiv\frac{v_{n+1,m}-v_{n-1,m}}{v_{n+1,m}+v_{n-1,m}}
\     \ ,\     \
b_{n,m}\equiv\frac{v_{n,m+1}-v_{n,m-1}}{v_{n,m+1}+v_{n,m-1}}
 \   , \end{equation} 
which are related to the variables $x$ and $y$ via 
\begin{equation} \label{eq:axy}  
a_{n,m}=\frac{y_{n,m}-x_{n-1,m}}{y_{n,m}+x_{n-1,m}}\    \ ,\    \  
b_{n,m}=\frac{1-x_{n,m-1}y_{n,m}}{1+x_{n,m-1}y_{n,m}}\   . 
\end{equation} 
By using the P$\triangle$E (\ref{eq:dMKdV}), we then get
\bse\label{eq:abxy}\bea
(1+rx_{n,m})(1+a_{n,m+1}) &=& 2-(1-r/y_{n,m})(1+a_{n,m})\   , 
\label{eq:abxya}\\ 
(x_{n,m}+r)(1+b_{n+1,m}) &=& 2r-(r-y_{n,m})(1+b_{n,m})\   ,  
\label{eq:abxyb}
\eea\ese 
The ODE (\ref{eq:altPVI}) (equivalent to \pvi ) 
is obtained from the differential-difference equation 
(\ref{eq:mkdvdif}) and the second relation (\ref{eq:abxyb}) together 
with the similarity constraint (\ref{eq:MKdVconstr}). More details can be 
found in Ref. \cite{NRGO}. 

Finally, to obtain the discrete and continuous modified \pvi\ 
equations in terms of the variable $h$ 
we solve the system of constraints (\ref{eq:dp34constr1}),  
(\ref{eq:dp34constr2}) as follows
\bse\label{eq:hhkk}\bea
\frac{2n}{h_{n-1,m}+h_{n,m}} &=& \frac{1}{h_{n,m}-k_{n,m}}
\left( \frac{1}{2}+\nu+n+m-\bmu k_{n,m}\right)\  , 
\label{eq:hhkka}\\
\frac{2m}{k_{n,m-1}+k_{n,m}} &=& \frac{1}{k_{n,m}-h_{n,m}}
\left( \frac{1}{2}+\nu+n+m-\bmu h_{n,m}\right)\  . 
\label{eq:hhkkb}
\eea\ese
These relations allow us to express $k$ entirely in 
terms of $h$ via 
\begin{equation}  \label{eq:k} 
k=\frac{\frac{1}{2}+\nu+n+m-\frac{2nh_{n,m}}{h_{n-1,m}+h_{n,m}}}
{\bmu-\frac{2n}{h_{n-1,m}+h_{n,m}}}\   , 
\end{equation} 
involving only shifts in the independent variable $n$. Furthermore, 
using the coupled P$\triangle$E's (\ref{eq:dKdV}) we obtain 
\begin{equation}  \label{eq:hhk} 
(h_{n,m}+k_{n+1,m})(h_{n,m}-k_{n,m})=(1-r^2)(h_{n,m}^2-1) \ , 
\end{equation} 
which is quadratic in the variable $h$. Inserting  (\ref{eq:k}) 
into (\ref{eq:hhk}) we obtain (\ref{eq:dP34}). Alternatively, we 
obtain a scond-order second-degree P$\Delta$E for $k$ by solving 
$h$ from (\ref{eq:hhk}) and inserting it into (\ref{eq:dKdV}). 

Another relation we need is the $m$-shifted object 
\begin{equation}\label{eq:hkk}
\frac{2n}{h_{n,m+1}+h_{n-1,m+1}}=\frac{r^2(k_{n,m}-h_{n,m})
[\frac{1}{2}+\nu+m-n-\bmu k_{n,m}]}{(1-r^2)
(k_{n,m}^2-1)}\   , 
\end{equation}
which follows from the $m$-shifted constraints (\ref{eq:dp34constr1}) 
and (\ref{eq:dp34constr2}) together with (\ref{eq:dKdV}).        
Similarly, we have 
\begin{equation}\label{eq:kk} 
\frac{2(n+1)}{h_{n+1,m}+h_{n,m}}=\frac{\frac{1}{2}+\nu+n-m+
\bmu k_{n+1,m}}{h_{n,m}+k_{n+1,m}}\   , 
\end{equation}
where $k_{n+1,m}$ can be eliminated using (\ref{eq:dKdV}).  

Finally, the continuous Schwarzian \pvi\ equation (\ref{eq:SPVI}) 
can be obtained from the coupled system of ODEs for $h$ and $k$ 
(\ref{eq:MPVIa}) and (\ref{eq:MPVIb}) using the 
differential-difference relation (\ref{eq:skdvdif}) which leads to 
\begin{equation}\label{eq:zh} 
 -t\frac{\pl}{\pl t}\log z_n=\frac{2n}{h_n+h_{n-1}}\   ,   
\end{equation} 
and consequently
\begin{equation}\label{eq:kh}
k-h=\frac{\frac{1}{2}+\nu+n+m-\bmu h}{\bmu     
+tz'/z}\   . 
\end{equation} 
Furthermore, the Miura relations (\ref{eq:cMiu}) and (\ref{eq:cinvMiu}) 
between \pvi\ in terms of $w$ and the M\pvi\ equation in terms of $h$ 
follow from the relation  
\begin{equation}  \label{eq:wh}
w(t)=t\frac{1-k}{1-h}
\end{equation}
where one can either eliminate $k$ by solving $h-k$ from 
(\ref{eq:MPVIb}) or differentiate (\ref{eq:wh}) and eliminate k by using 
(\ref{eq:wh}) and both (\ref{eq:MPVIa}) and (\ref{eq:MPVIb}). 
The discrete Miura transformations (\ref{eq:xh}) and (\ref{eq:hx}) 
follow from the definitions (\ref{eq:xy}) and (\ref{eq:hk}) exploiting the 
various relations such as (\ref{eq:kk}) and (\ref{eq:kh}) above. 
The consistency of all these connections and relations follow from 
the structures given in \cite{NRGO} from which also isomonodromic deformation 
systems (i.e. Lax pairs) for all the equations can be derived as well as 
explicit gauge transformations between them. The construction of those is 
in preparation. 
 
\section{Some Special Limits}
\setcounter{equation}{0}
 
In this section, we consider limits of the equations presented in \S 2. 
In particular, we present new limits exhibiting discrete versions 
of Schwarzian \pii\ (\ref{eq:SPII}). We want to stress again, 
that the discrete and continuous equations of the previous sections 
are not related via continuum limit (which amount, in fact, to coalescence limits on the parameters of the equations), but that their relation is exact 
in the sense that the discrete and continuous flows are compatible, meaning 
that one can be considered to be a symmetry of the other and vice versa.  
However, in this section we will perform limits on the parameters 
to show that the discrete equations contain many of the previously known 
discrete Painlev\'e equations under coalescence, in particular the discrete 
analogues of \pii . In fact, taking the limit
$r\rightarrow\infty$, $m/r\rightarrow\xi$, it is easy to see that
(\ref{eq:mpii}) reduces to
\begin{equation} \label{eq:altdpii}
\frac{n+1}{x_nx_{n+1}+1}+\frac{n}{x_nx_{n-1}+1} =
n+\bmu+\frac{1}{2}\xi\left(x_n-\frac{1}{x_n}\right)\   ,
\end{equation} 
which is the the alternate d\pii\ equation of \cite{FGR,altdp2}. 
 
The limit to the standard discrete \pii\ (d\pii ) 
is a little more subtle, since we need
to work in an oblique direction. In fact, to obtain d\pii\ from
(\ref{eq:mpii}) we write $r=1+\dd$ and take the limit $\dd\rightarrow
0$, whilst taking $n=n^\prime-m$, where $n,m\rightarrow\infty$
such that $n^\prime$ is fixed (in the limit) and $\dd m\rightarrow
\eta$ finite. In that limit we have also $(1-x_n)/(1+x_n)\rightarrow
a_{n+1}$ and (\ref{eq:mpii}) reduces to the following equation:
\begin{equation} \label{eq:limit}
(n^\prime+1)(1-x_n^2)-(\mu+\nu)(1+x_n)^2 =
2\eta x_n\left(\frac{1-x_{n+1}}{1+x_{n+1}}+\frac{1-x_{n-1}}
{1+x_{n-1}}\right)\   ,
\end{equation} 
where we consider $x_n$ to be a function of $n^\prime$ rather
than $n$. In terms of $a_n$, (omitting from now on primes
on the $n^\prime$ variable), this equation reads
\begin{equation} \label{eq:dpii}
\frac{1}{2}\eta(a_{n+1}+a_{n-1})=\frac{-\mu+\nu+na_n}{1-a_n^2}\  ,
\end{equation} 
which is d\pii\ \cite{NP,Per}. 

In the above coalescence limits, the discrete Schwarzian equation 
(\ref{eq:dSPVI}) reduces to 
\begin{equation} \label{eq:sdpii}
4\eta\frac{(z_{n+2}-z_{n+1})(z_n-z_{n-1})}{(z_{n+2}-z_n)
(z_{n+1}-z_{n-1})}= \frac{1}{2}+\nu+n-
\bmu\frac{z_{n+1}+z_n}{z_{n+1}-z_n}\    , 
\end{equation} 
which is the discrete Schwarzian \pii , i.e. the discrete 
analogue of (\ref{eq:SPII}), cf. \cite{Dorf}, whereas in the alternate 
limit the equation becomes 
\begin{equation}\label{eq:alt-sdpii}
\left[\bmu\frac{z_{n+1}}{z_{n+1}-z_n}-
(n+1)\frac{z_{n+2}-z_{n+1}}{z_{n+2}-z_n}\right] 
\left[\bmu\frac{z_n}{z_{n+1}-z_n}-
n\frac{z_n-z_{n-1}}{z_{n+1}-z_{n-1}}\right] = \xi^2\   , 
\end{equation}  
cf. \cite{Carg}. In both equations the ``discrete Schwarzian 
derivative'', which is the canonical cross-ratio of the four 
entries $z_{n-1}$, $z_n$, $z_{n+1}$, $z_{n+2}$ is prominent.  
Finally, in these limits the modified equations reduce to 
\begin{equation} \label{eq:dp34} 
(h_{n+1}+h_n)(h_n+h_{n-1})=4\eta\frac{h_n^2-1}{\bmu 
h_n-(\frac{1}{2}+\nu+n)}\   , 
\end{equation} 
which is the discrete analogue of 
the thirty-fourth equation in the classification list of
Painlev\'e-Gambier (see \cite{ince}).
Its discrete analogue d\ptf, i.e. (\ref{eq:dp34}), was first given in
\cite{RG} (see also \cite{JRG}). The alternate version, following 
from the other coalescence limit, reads 
\begin{equation} \label{eq:alt-dp34} 
\left(\bmu-\frac{2(n+1)}{h_{n+1}+h_n}\right) 
\left(\bmu-\frac{2n}{h_{n-1}+h_n}\right) =
\frac{\xi^2}{h_n^2-1}\   , 
\end{equation} 
and was first given in \cite{Carg}. 

\subsection*{Acknowledgements}

FWN would like to thank the Department of Pure Mathematics of the 
University of Adelaide for its hospitality during a visit where 
the present work was performed. The work was supported by the 
Australian Research Council. The authors would also like to thank 
C. Cosgrove for clarifying a point of detail concerning the 
Painlev\'e-Gambier classification.

\end{document}